\documentclass[a4paper,twocolumn,11pt,accepted=2024-01-11]{quantumarticle}
\pdfoutput=1
\usepackage[utf8]{inputenc}
\usepackage[english]{babel}
\usepackage[T1]{fontenc}
\usepackage{amsmath}
\usepackage{hyperref}
\usepackage{tikz}
\usepackage{lipsum}
\usepackage{amssymb,amsmath}
\usepackage{mathrsfs}
\usepackage{graphicx}
\usepackage{color}
\usepackage{hyperref}
\usepackage{float}
\usepackage{listings}
\usepackage{tikz}
\usepackage{qcircuit}
\usepackage{algorithm}  
\usepackage{algpseudocode}  
\usepackage{booktabs}
\usepackage{multirow}
\usepackage{threeparttable}
\usepackage{makecell}
\usepackage{subfigure}
\usepackage{hyperref}

\begin{document}

\title{A new quantum machine learning algorithm: split hidden quantum Markov model inspired by quantum conditional master equation}

%author1
\author{Xiao-Yu Li}
\affiliation{School of Information and Software Engineering, University of Electronic Science and Technology of China, Cheng Du, 610054, China}
% \orcid{0000-0002-2445-2701}
% \email{latex@quantum-journal.org}
% \homepage{http://quantum-journal.org}

%author2
\author{Qin-Sheng Zhu$^\dagger$}
\affiliation{School of Physics, University of Electronic Science and Technology of China, Cheng Du, 610054, China}
\affiliation{Advanced Cryptography and System Security Key Laboratory of Sichuan Province, Cheng Du, 610103, China}
\thanks{Corresponding:zhuqinsheng@uestc.edu.cn}

%author3
\author{Yong Hu}
\affiliation{School of Physics, University of Electronic Science and Technology of China, Cheng Du, 610054, China}

%author4
\author{Hao Wu}
\affiliation{School of Physics, University of Electronic Science and Technology of China, Cheng Du, 610054, China}
\affiliation{Institute of Electronics and Information Industry Technology of Kash, Kash, 844000, China}

%author5
\author{Guo-Wu Yang}
\affiliation{School of Computer Science and Engineering, University of Electronic Science and Technology of China, Cheng Du, 610054, China}

%author6
\author{Lian-Hui Yu}
\affiliation{School of Physics, University of Electronic Science and Technology of China, Cheng Du, 610054, China}

%author7
\author{Geng Chen}
\affiliation{School of Computer Science and Engineering, University of Electronic Science and Technology of China, Cheng Du, 610054, China}
\maketitle

%\pagewiselinenumbers
%\switchlinenumbers

\begin{abstract}
The Hidden Quantum Markov Model (HQMM) has significant potential for analyzing time-series data and studying stochastic processes in the quantum domain as an upgrading option with potential advantages over classical Markov models. In this paper, we introduced the split HQMM (SHQMM) for implementing the hidden quantum Markov process, utilizing the conditional master equation with a fine balance condition to demonstrate the interconnections among the internal states of the quantum system. The experimental results suggest that our model outperforms previous models in terms of scope of applications and robustness. Additionally, we establish a new learning algorithm to solve parameters in HQMM by relating the quantum conditional master equation to the HQMM. Finally, our study provides clear evidence that the quantum transport system can be considered a physical representation of HQMM. The SHQMM with accompanying algorithms present a novel method to analyze quantum systems and time series grounded in physical implementation.
\end{abstract}

\section{Introduction} 

The significant increase of data and information has emphasized that classical algorithms have difficulties in meeting computation demands for efficiency and speed. Therefore, quantum computing, capable of efficient computation, has become a viable solution. Unlike classical computing, quantum computing employs quantum bits' superposition for data storage, reading, and Efficient computing. As a result, quantum computing promises to solve problems that are too complex for classical computers.

Quantum computing has made great improvements in hardware implementations \cite{Cirac, Knill}  and algorithmic approaches \cite{Jacob, cerezo2022challenges}, propelled the field of quantum computing into the \textit{"Noisy intermediate-scale quantum (NISQ) algorithms"} era in the last decade. During that era, many hybrid framework \cite{Kishor} algorithms emerged that combined quantum and classical approaches to cope with suboptimal hardware conditions. Initial applications of quantum computing in chemistry \cite{Aspuru}, Hamiltonian simulation \cite{Georgescu}, biology \cite{Reiher}, pharmaceutical \cite{Cao}, finance \cite {Orus}, materials \cite{Dallaire} and various other fields demonstrate its potential advantages over classical computing.

In the field of machine learning, the Hidden Markov Model (HMM) algorithm plays a vitall role and has been extensively and effectively utilized in domains such as stock market forecasting  \cite{fons2021a,chandrika2020application}, natural language processing \cite{suleiman2017the,muhammad2018speech}, protein sequencing \cite{sonnhammer1998a,xie2021shortest}.

The classical HMM \cite{eddy2004hidden} has three main parts: training, decoding, and learning. When the dimensionality of the hidden state is not large, the Baum-Welch algorithm \cite{baggenstoss2001modified}, the Viterbi algorithm \cite{kavcic2000viterbi}, and the EM algorithm \cite{moon1996expectation} can be used to solve these problems efficiently. However, when the dimensions of the hidden state and the observation space increase simultaneously, the solution speed and accuracy of the classical algorithms become weak.

Quantum algorithms were introduced in response to address such problems and find more effective solutions. Monras \textit{et.al.}    \cite{Alex} gave the mathematical definition of the hidden quantum Markov process in terms of Kraus operators. This key work offers insight into  HQMM and a method for studying this model from the perspective of quantum open systems, as well as creating heuristic quantum computing algorithms. Compared to the classical HMM, the quantum version also functions as a stochastic probability graph model, and it includes the same three main parts. Crucially, the advantages of HQMM are gradually becoming apparent with the study of problem solving the model parameters.

Srinivasan \textit{et. al.}\cite{Siddarth} proposed an algorithm based on the Norm Observable Operator Model's learning algorithm, originally presented by Jaeger \textit{et.al.}\cite{Jaeger2000Observable}. This study demonstrates that HQMM offers advantages over traditional algorithms with regard to model complexity and accuracy. Nevertheless, this model is only suitable for situations where the hidden state dimension is relatively small, and it is inclined to succumb to local optimal solutions. Liu \textit{et.al.} \cite{LiuQing} analytically demonstrated the superiority of the quantum stochastic model in comparison to the classical model, providing qualitative evidence of the quantum advantage that HQMM possesses. In further work \cite{elliott2021memory}, they also highlight that the quantum implementation of the HMM could both mitigate thermal dissipation and achieve an advantage in memory compression. In 2020, Adhikary \textit{et.al.} \cite{Sandesh} proposed a learning algorithm grounded in optimization theory on the manifold \cite{Bojiang} with goal of solving the Kraus, which provides an algorithmic basis for substantial applications of HQMM. The work of Markov \textit{et.al.} \cite{markov2022implementation} in 2022 shows that HQMM has a significant advantage in state space complexity over stochastic process languages. In 2023, Li \textit{et.al.} \cite{li2022simulating} presents a new algorithm for modeling the dynamics of Markovian open quantum systems that is cleaner and more efficient than previous algorithms. 

Taken together, the above work show that HQMM has potential advantages that need to be explored. The primary motivation of this paper is to further explore HQMM from a physical perspective, providing new tools for a nuanced understanding of the intricate states of quantum systems while ensuring algorithmic performance.
 
 We developed a new HQMM learning algorithm, namely SHQMM, using the quantum conditional master equation on an open quantum system described by the quantum master equation \cite{Yoshitaka,Ishizaki,Jin} building upon the work of Clark et al. \cite{Clark}. As a result, SHQMM has the capability to function with open systems and non-unitary quantum algorithms, rendering it applicable to the noise-related issues that arise in quantum computers during NISQ. Another aspect, SHQMM provides an understanding of the quantum state of the system under fine balance condition while guaranteeing algorithmic performance, provides HQMM with interpretability that corresponds to quantum transport systems \cite{Xin2005}. Additionally, our work is expected to provide new ideas for the physical implementation of quantum neural network (QNN).

The contribution of this paper is as follows:

1. A new SHQMM learning algorithm is constructed on open quantum systems by using the quantum conditional master equation, which is able to handle open systems and non-unitary quantum algorithms.

2. SHQMM with the introduction of periodic boundary conditions guarantees the performance of the algorithm under fine balance conditions, which can be physically related to quantum transport systems. Model possesses a clear physical representation, well-defined dimensional relationships, and robust interpretability.

3. Numerical experiments show that SHQMM achieves better results on both quantum and classical data sets and is robust to random initialization. Exploration of the quantum system did not degrade the performance of the model.

The paper is organized as follows: Sec. \ref{Related work} provides an introduction of related work, including the basic concepts of HMM, HQMM and quantum conditional master equation. In Sec. \ref{The split hidden quantum Markov model(SHQMM) based on QCME}, we formally introduce the SHQMM, a novel stochastic probabilistic graphical model. The implementation of SHQMM is demonstrated using a quantum transport system as an illustrative example.  Sec. \ref{Experiment and Results} presents the results of numerical experiments on the SHQMM model, demonstrating its adaptability on both quantum and classical data, robust to random initialization, and performance on DA. In Sec. \ref{discussion}, we compare SHQMM to previous work, analyze its complexity, and discuss the benefits. Finally, in Sec. \ref{conclusion}, we summarize our work.

\section{Related work \label{Related work}}
\subsection{Hidden quantum Markov model \label{Hidden quantum Markov model}}
The Hidden Markov Model (HMM) is a type of probabilistic graph model that describes the evolutionary properties of Markov dynamics. It consists of two important parameters: the transition matrix $\mathbf{T}$ and the observation matrix $\mathbf{C}$, which are constant matrices. A HMM can be defined as $\lambda=(\mathbf{T},\mathbf{C},x_{0})$, where $x_{0}$ is the initial state vector. The update of the hidden state and the observable results can be obtained from Eq.\ref{HMM}, which represents a state-emitting (Moore) hidden Markov model.

\begin{equation}\label{HMM}
	\begin{aligned}
		&x_{t+\Delta t}=\mathbf{T}x_{t}\\
		&y_{t+\Delta t}={\rm diag}(\mathbf{C}_{(y,:)})x_{t+\Delta t} ,
	\end{aligned}
\end{equation}

where the variable $y$ represents an output symbol in the observable space $O$.

A Hierarchical Quantum Markov Model (HQMM) can be defined using a set of parameters $\lambda_{Q}=(\rho_{0},{K_{y}})$, similar to the classical Markov process. Here, $\rho_{0}$ corresponds to the initial state vector $x_0$ of the classical Markov model, and the Kraus operator $K_{y}$ corresponds to the matrices $\mathbf{T}$ and $\mathbf{C}$. In comparison to the classical Markov model, the Kraus operator ${K_{y}}$ plays a dual role as both the evolution state and the observable output result. It satisfies the condition $\sum_{m}K_{m}^{\dagger}K_{m}=I$. When the system is measured (assuming the measurement or read-out result is $y$), the density matrix can be expressed as follows \cite{Sandesh}: 

\begin{equation}\label{eq5}
	\rho_{y}(t+\Delta t)=\frac{\sum_{\omega_{y}}K_{\omega_{y}}\rho(t)K_{\omega_{y}}^{\dagger}}{{\rm Tr}[\sum_{\omega_{y}}K_{\omega_{y}}\rho(t)K_{\omega_{y}}^{\dagger}]}
\end{equation}

where $\omega_{y}$ denotes the auxiliary dimension of Kraus operator.

The difference between the HMM and the HQMM is shown in Table.\ref{difference}. 
\begin{table}[h]
	\caption{The difference between the HMM and the HQMM}
	\renewcommand{\arraystretch}{1.3}
	\centering
	\begin{small}
		\begin{tabular}{|c|c|c|}
			\hline
			Model&  HMM & HQMM \\
			\hline
			State& state vector $x$ & density matrix $\rho$ \\
			\hline
			\makecell[c]{Transition\\and Emission}  & matrix $\mathbf{T}, \mathbf{C}$ & \makecell[c]{Kraus operators\\$\{K\}$} \\
			\hline
			\makecell[c]{Steady\\State} & $x^{*}=\mathbf{T}x^{*}$ & $\rho^{*}=\sum_{\omega_{y}}{K}_{\omega_{y}}\rho^{*}{K}_{\omega_{y}}^{\dagger}$\\
			\hline
			Probability & $\mathbb{I}{\rm diag}(\mathbf{C}_{(y,:)})x$ & ${\rm Tr}(\sum_{\omega_{y}}{K}_{\omega_{y}}\rho {K}_{\omega_{y}}^{\dagger})$\\
			\hline
		\end{tabular}
		\label{difference}
	\end{small}
\end{table}

To calculate the parameters $\{K\}$ of the HQMM, Adhikary \textit{et.al.} \cite{Sandesh} proposed a maximum likelihood estimation algorithm. This algorithm assumes that a set of observation sequences $y_{1},y_{2},y_{3},\cdots,y_{T}$ is known, and constructs the maximum likelihood function based on this data. This is a particular case where $\omega=1$:

\begin{equation}
	\mathscr{L}=-{\rm ln}\, {\rm tr}\left(K_{y_{T}}\cdots K_{y_{2}}K_{y_{1}}\rho_{0}K_{y_{1}}^{\dagger}K_{y_{2}}^{\dagger}\cdots K^{\dagger}_{y_{T}}\right).
\end{equation}

Then the parameter solving problem of the HQMM is transformed into a constrained optimization problem:

\begin{equation}\label{eq7}
	\begin{aligned}
		&{\rm minmize}_{\{K\}}\quad \mathscr{L}(\{K\})\\
		&{\rm subject \ to}\quad \sum_{y}K_{y}^{\dagger}K_{y}=I,K_{y} \in \mathbb{C}^{n\times n}.
	\end{aligned}
\end{equation}

Stack $K_{m}$ by column to form a new matrix $\kappa=[K_{1},K_{2},\cdots,K_{m}]^{ T}$ with dimension $nm\times n$, the constraint condition in Eq.\ref{eq7} can be rewritten as
\begin{equation}\label{eq8}
	\kappa^{\dagger}\kappa=I,\kappa\in \mathbb{C}^{nm\times n}.
\end{equation}
According to Ref.\cite{Sandesh}, $\kappa$ in Eq.\ref{eq8} lies on the Stiefel manifold, and the following gradient descent method can be used to solve Eq.\ref{eq7}:

\begin{equation}\label{eq9}
	\begin{aligned}
		&G= \frac{\partial \mathscr{L}}{\partial \kappa},\\
		&\kappa=\kappa-\tau \mathbf{U} (I+\frac{\tau}{2}\mathbf{V}^{\dagger}\mathbf{U})^{-1}\mathbf{V}^{\dagger}\kappa.
	\end{aligned}
\end{equation}

In Eq.\ref{eq9}, $\mathbf{U}=[G,\kappa]$, $\mathbf{V}=[\kappa,-G]$, and $\tau$ is a positive real number.

\subsection{The quantum conditional master equation\label{quantum conditional master equation} }
As a general type of equation describing the evolution of open quantum systems, the quantum conditional master equation adopts the specific form of the Lindbladian equation \cite{kastoryano2023quantum}, with particular consideration given to the interaction with the environment.

For an open quantum system, the Hamiltonian form can be applied as

\begin{equation}\label{form Hamilatonian}
	H = H_{S} + H_{E} + H^{\prime}.
\end{equation}

$H_S$ and $H_{E}$ represent the Hamiltonians of the quantum system and the environment, respectively. The Hamiltonian $H^{\prime}$ describes the coupling effect between the quantum system and the environment. In the case of weak coupling between the quantum system and the environment, $H^{\prime}$ can be treated as a perturbation. Using the expansion of the second-order cumulant, we can obtain a description of the evolution of the reduced density matrix. 

\begin{equation}\label{qme}
	\begin{aligned}
		\dot{\rho}(t)&= -i\mathcal{L}\rho(t)\\&-\int^{t}_{0}d\tau\langle \mathcal{L}^{\prime}(t)\mathcal{G}(t,\tau)\mathcal{L}^{\prime}(\tau)\mathcal{G}^{\dagger}(t,\tau)\rangle\rho(t).
	\end{aligned}
\end{equation}  

Here, the Liouvillian super operator is defined as $\mathcal{L}=[H_{S},(\cdots)]$, $\mathcal{L}^{\prime}=[H^{\prime}_{S},(\cdots)]$. $\mathcal{G}(t,\tau)=G(t,\tau)\times(\cdots)\times G^{\dagger}(t,\tau)$. $G(t,\tau)$ is the Green's function related to $H_S$. The reduced density matrix is obtained by partially tracing the density matrix of the composite system, that is, $\rho(t)={\rm Tr}_{E}[\rho_T(t)]$. 

\begin{figure}[h]
	\centering
	\includegraphics[scale=0.55]{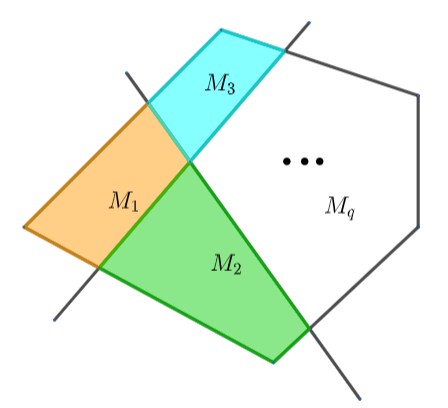}
	\caption{The diagram of dividing the environment space. The Pentagon represents the Hilbert space of the environment, which is divided into some subspace $M_{1}$, $M_{2}$, $M_{3}$, $\cdots$, $M_{q}$ corresponding to $\rho^{(\mathcal{M}_1)},\rho^{(\mathcal{M}_2)} \cdots, \rho^{(\mathcal{M}_q)}$.}
	\label{split_space}
\end{figure}

In experiments, measurement results are typically linked to changes in the internal state of a system. Therefore, unlike the method used to derive Eq.\ref{qme}, Li et. al. \cite{Xin2005} introduced the "detailed balance" condition to illustrate the relationship among different system states when studying the current ( the measurement result) of a quantum transport system. This allowed them to derive the quantum conditional master equation (QCME) in Equation \ref{qcme} and to obtain some interesting results. By adopting the approach of the QCME described in reference \cite{Xin2005} and considering the detailed balance among different system states, the general QCME can be expressed as following when the environment space is divided into different subspaces as shown in Figure \ref{split_space}.

\begin{equation}\label{form qcme}
		\begin{aligned}
		\dot{\rho}^{(\mathcal{M}_{q})}&=-i\mathcal{L}\rho^{(\mathcal{M}_{q})}\\
		&-\int^{t}_{0}d\tau{\rm Tr}_{E^{(\mathcal{M}_{q})}}[\mathcal{L}^{\prime}(t)\mathcal{G}(t,\tau)\mathcal{L}^{\prime}(\tau)\mathcal{G}^{\dagger}(t,\tau)\rho_{T}(t)].
		\end{aligned}
\end{equation}

 Here, the proposed initial conditions for the quantum conditional master equation are $\rho_{T}(0) \simeq \sum_{\mathcal{M}_{q}}\rho^{(\mathcal{M}_q)}(0) \otimes \rho_{E}^{(\mathcal{M}_q)}(0) $, and $\rho^{(\mathcal{M}_{q})}$ denotes \emph{conditional density matrix} of the quantum system corresponding to the environment $\rho^{(\mathcal{M}_{q})}_{E}$ associated with the subspace $M_{q}$. Note that  $\rho^{(\mathcal{M}_{q})}$ also satisfies positive semi-definite, and ${\rm Tr}(\rho^{(\mathcal{M}_{q})})\leq 1$, ${\rm Tr}[\sum_{\mathcal{M}_{q}}\rho^{(\mathcal{M}_{q})}]=1$.
The use of quantum conditional master equation enables the representation of the relationship between the different subspaces $M_{q}$, which provides a better understanding of the open quantum system being studied. 

\section{The SHQMM based on QCME \label{The split hidden quantum Markov model(SHQMM) based on QCME}}

\subsection{The quantum master equation of quantum transport system \label{The quantum master equation of quantum transport system} }

\textit{Since any quantum computing needs an actual physical system to implement, we need search for an open quantum system that can be described by the conditional master equation to establish HQMM. We found that the quantum transport system is suited for implementing our HQMM based on previous work \cite{Sandesh, Xin2005}.}

As a result, in this section, we present the quantum conditional master equation for the quantum transport system. The Hamiltonian of this quantum system is expressed as follows \cite{Xin2005}:

\begin{equation}\label{Hamiltonian}
	\begin{aligned}
		H&=H_{S}(a_{\mu}^{\dagger},a_{\mu})+\sum_{\alpha=L,R}\sum_{\mu k}\epsilon_{\alpha\mu k}d^{\dagger}_{\alpha \mu k}d_{\alpha \mu k}\\
		&+\sum_{\alpha=L,R}\sum_{\mu k}(t_{\alpha \mu k}a^{\dagger}_{\mu}d_{\alpha\mu k}+{\rm H.c}).
	\end{aligned}
\end{equation}

where $H_{S}$ is the Hamiltonian of the quantum dots system, $L$ and $R$ represent the left and right electrodes respectively, $d^{\dagger}_{\alpha\mu k}$ and $d_{\alpha \mu k}$  represent the creation and annihilation operators of electrons in the electrode, respectively, and $t_{\alpha\mu k}$ represents the coupling strength between the electrode and the quantum dot system. The master equation for the quantum transport system can be derived through some calculations \cite{Xin2005} based on Eq.\ref{qme}:

\begin{equation}\label{qme_qts}
	\dot{\rho} = -i\mathcal{L}\rho-\frac{1}{2}\sum_{\mu}\{[a_{\mu}^{\dagger}, A_{\mu}^{(-)}\rho-\rho A^{(+)}_{\mu}]+{\rm H.c.}\}.
\end{equation}

 If the state space where the electrode is located, without any electrons passing through the quantum dot system, is denoted as $E^{(0)}$, it is formed by the wave function of the two isolated electrodes on the left and right $E^{(0)}={\rm span}\{|\psi_{L}\rangle\otimes|\psi_{R}\rangle\}$. If there are $n$ electrons from
the state space where the right electrode passes through the quantum dot to the left electrode, it is denoted as $E^{(n)}$ ($n=1,2,3,\cdots$).
Then the electrode state space $E$ in the equation can be decomposed as $E=\oplus_{n}E^{(n)}$, which leads to the quantum conditional master equation \cite{Xin2005} with an initial condition of $\rho_{T}(0)\simeq\sum_{n}\rho^{(n)}(0)\otimes \rho_{E}^{(n)}(0)$.

\begin{equation}\label{qcme}
	\begin{aligned}
		\dot{\rho}^{(n)}&=-i\mathcal{L}\rho^{(n)}-\frac{1}{2}\sum_{\mu}\{[a^{\dagger}_{\mu}A^{(-)}_{\mu}\rho^{(n)}+\rho^{(n)}A_{\mu}^{(+)}a_{\mu}^{\dagger}\\&-A_{L\mu}^{(-)}\rho^{(n)}a_{\mu}^{\dagger}
		-a^{\dagger}_{\mu}\rho^{(n)}A_{L\mu}^{(+)}-A^{(-)}_{R\mu}\rho^{(n-1)}a_{\mu}^{\dagger}\\&-a_{\mu}^{\dagger}\rho^{(n+1)}A_{R\mu}^{(+)}]+{\rm H.c.}\}.
	\end{aligned}
\end{equation}

Here, $\rho^{(n)}={\rm Tr\_}_{E^{(n)}}[\rho_{T}(t)]$ is the conditional density matrix of the quantum dot system \cite{Xin2005}, which means that there are $n$ electrons passing through the quantum dot system within time $t$. Here, the number of electrons $n$ corresponds to the subspace $\mathcal{M}_{q}$ in Eq. \ref{form qcme}.

\subsection{The relationship between the QCME equation and HQMM \label{The Relationship Between QCME and HQMM}}

 Based on the contents in Sec.\ref{quantum conditional master equation} and \ref{The quantum master equation of quantum transport system}, we derive the hidden quantum Markov model from a quantum master equation and propose a novel stochastic graph model from a quantum conditional master equation. After some calculations (the detailed proof and calculation process are shown in supplemental material, we obtain:

(1) For quantum master equation of Eq.\ref{qme}, the evolution density matrix of quantum dot system is

\begin{equation}\label{hqmm_qme}
	\rho(t+\Delta t)=\sum_{i,\mu}K_{i,\mu}\rho K_{i,\mu}^{\dagger}.
\end{equation}

(2) For quantum conditional master equation of Eq.\ref{qcme},  the evolution density matrix of quantum dot system is

\begin{equation}\label{eq14}
	\begin{aligned}
		\rho^{(n)}(t+\Delta t)&=\sum_{i,\mu}K_{i,\mu}\rho^{(n)}K_{i,\mu}^{\dagger}\\
		&+ \sum_{\mu}K_{3,\mu}\rho^{(n-1)}K_{3,\mu}^{\dagger}\\
		&+ \sum_{\mu}K_{4,\mu}\rho^{(n+1)}K_{4,\mu}^{\dagger},
	\end{aligned}
\end{equation}

where $i=0,1,2$.  

Comparing Eq.\ref{eq5}($\omega_y=1$) and Eq.\ref{hqmm_qme}, we concluded that there is a close relationship between a quantum Markov model and a quantum master equation. However, the Kraus operators $K_{i,\mu}$ of Eq.\ref{eq14} are involved with the related $\rho^{(n)}$, $\rho^{(n-1)}$, and $\rho^{(n+1)}$. This difference arises from the division of the Hilbert space of the environment, which gives rise to a new HQMM called the split hidden quantum Markov model.

\subsection{Split hidden quantum Markov model \label{split hidden quantum Markov model}}

In this section, we present the SHQMM inspired by quantum transport systems, which is the main theoretical improvement in this work. Similar to the HQMM, the SHQMM is defined by applying a set of parameters $\lambda_{SQ}=({\rho^{(0)}, \rho^{(1)}, \rho^{(2)}, \cdots}, {K_{y}})$ where ${\rm Tr}(\sum_{i=0} \rho^{(i)})=1$.

Firstly, the evolution conditional density matrix of quantum system $H_{S}$ is written as

\begin{equation}\label{Form1}
	\begin{aligned}
		\rho^{(\mathcal{M}_{q})}(t+\Delta t)&= \sum_{y} K_{y}^{(\mathcal{M}_{1})}\rho^{(\mathcal{M}_{1})}(t)K_{y}^{(\mathcal{M}_{1})\dagger} + \cdots \\&+ \sum_{y} K_{y}^{(\mathcal{M}_{q})}\rho^{(\mathcal{M}_{q})}(t)K_{y}^{(\mathcal{M}_{q})\dagger} + \cdots,
	\end{aligned}
\end{equation}

where, $q$ denotes the values of subspace for environment $\bf{M}$ and $\sum_{i,\mu}K_{i,\mu}^{\dagger}K_{i,\mu}=I$. The parameter $y$ represents the read-out of information symbols from the open quantum system. Eq.\ref{Form1} represents a comprehensive expression that correlates to the relationship between $\rho^{(\mathcal{M}_{q} )} $.

Secondly, when we read out or measure a certain value $y'$ for $\rho^{(\mathcal{M}_{q})}(t)$, the conditional density matrix $\rho^{(\mathcal{M}_{q})}(t+\Delta t)$ is rewritten as follows:

\begin{equation}\label{Gen}
	\begin{aligned}
		\rho_{y'}^{(\mathcal{M}_{q})}(t+\Delta t)&= \frac{\rho_{y'}^{'(\mathcal{M}_{q})}(t+\Delta t)}{Tr[\sum_{\mathcal{M}_{q}}{\rho^{'}}_{y'}^{(\mathcal{M}_{q})}(t+\Delta t)]},\\
		\rho_{y'}^{'(\mathcal{M}_{q})}(t+\Delta t)&=  K_{y'}^{(\mathcal{M}_{1})}\rho^{(\mathcal{M}_{1})}(t)K_{y'}^{(\mathcal{M}_{1})\dagger} + \cdots \\&+ K_{y'}^{(\mathcal{M}_{q})}\rho^{(\mathcal{M}_{q})}(t)K_{y'}^{(\mathcal{M}_{q})\dagger} + \cdots.
	\end{aligned}
\end{equation}

Thirdly, the probability of obtaining the measurement result $y'$ is given by:

\begin{equation}\label{Pro Gen}
		P(y')= \sum_{\mathcal{M}_{q}} Tr[\rho_{y'}^{'(\mathcal{M}_{q})}(t+\Delta t)].
\end{equation}

Here, Eq.\ref{Pro Gen} describes the contribution of different $\rho'^{(\mathcal{M}_{q})}$ to the probability $P(y')$, and this process reveals the concept of "detailed balance" in physics, as described in Eq.\ref{Form1}.\\

\paragraph{The concretely implement example of our SHQMM}

\begin{figure}[b]
	\centering
	\begin{tikzpicture}
		[inner sep=1mm, place/.style={circle,draw=black!60, fill=gray!50, thick},
		pre/.style={<-,shorten <=1pt,>=stealth,color=black,semithick},
		post/.style={<-,shorten <=1pt,>=stealth,color=red,semithick},
		cubic/.style={<-,shorten <=1pt,>=stealth,color=blue,semithick}]
		% \node[place] (1) at(0,2){};
		\node[place] (u1) [label=left :\footnotesize $\rho_{0}^{(0)}$]at(0,0) {};
		\node[place] (u2) [label=left:\footnotesize$\rho_{0}^{(1)}$]at(0,-1.2) {};
		\node[place] (u3) [label=left:\footnotesize $\rho_{0}^{(2)}$]at(0,-2.4) {};
		\node[place] (u4) [label=left:\footnotesize $\rho_{0}^{(3)}$]at(0,-3.6) {};
		\node  (dot) at (0,-4.2)  {\bf{$\vdots$}};
		\node[place] (u5) [label=left:\footnotesize $\rho_{0}^{(n)}$]at(0,-5) {};
		\node[place] (r1) []at(1.2,0) {}
		edge [post] (u1)
		edge [cubic] (u2);
		\node[place] (r2) []at(1.2,-1.2) {}
		edge [pre] (u1)
		edge [post] (u2)
		edge [cubic] (u3);
		\node[place] (r3) []at(1.2,-2.4) {}
		edge [pre] (u2)
		edge [post] (u3)
		edge [cubic] (u4);
		\node[place] (r4) []at(1.2,-3.6) {}
		edge [post] (u4)
		edge [pre]  (u3);
		\node  (dot) at (1.2,-4.2)  {\bf{$\vdots$}};
		\node[place] (r5) []at(1.2,-5) {}
		edge [post] (u5); 
		\node[place] (L1) [] at (2.4,0){}
		edge [post] (r1)
		edge [cubic] (r2);
		\node[place] (L2) []at (2.4,-1.2) {}
		edge[pre] (r1)
		edge[post] (r2)
		edge [cubic] (r3);
		\node[place] (L3) []at (2.4,-2.4) {}
		edge[pre] (r2)
		edge[post] (r3)
		edge [cubic] (r4);
		\node[place] (L4) []at (2.4,-3.6) {}
		edge [pre] (r3)
		edge [post] (r4);
		\node  (dot) at (2.4,-4.2)  {\bf{$\vdots$}};
		\node[place](L5) []at (2.4,-5) {}
		edge[post] (r5);
		\draw [dashed,cubic] (r4) -- (u5);
		\draw [dashed, pre]  (r5) -- (u4);
		\draw[dashed, cubic] (L4) -- (r5);
		\draw [dashed, pre]  (L5) -- (r4);
		\node (e1) at (4.5,0) [place] {};
		\node [place] (e2) [label=right:\footnotesize$\rho_{T}^{(1)}$] at (4.5,0-1.2){};
		\node [place] (e3) [label=right:\footnotesize$\rho_{T}^{(2)}$]at (4.5,-2.4) {};
		\node [place] (e4) [label=right:\footnotesize$\rho_{T}^{(3)}$]at (4.5,-3.6) {};
		\node (dot) at (4.5, -4.2) {\bf{$\vdots$}};
		\node[place] (e5) [label=right:\footnotesize $\rho_{T}^{(n)}$]at (4.5,-5) {};
		\node (dot) at (3.5,0) {\bf{$\cdots$}};
		\node (dot) at (3.5,-1.20) {\bf{$\cdots$}};
		\node (dot) at (3.5,-2.40) {\bf{$\cdots$}};
		\node (dot) at (3.5,-3.60) {\bf{$\cdots$}};
		\node (dot) at (3.5,-5.0) {\bf{$\cdots$}};
		\node [black, above] at (u1.north) {\footnotesize$t=0$};
		\node [black, above] at (r1.north) {\footnotesize$t=1$};
		\node [black, above] at (L1.north) {\footnotesize$t=2$};
		\node [black, above] at (e1.north) {\footnotesize$t=T$};
		\node [black, right] at (e1.east) {\footnotesize $\rho_{T}^{(0)}$};
	\end{tikzpicture}
	\caption{The expanded calculation diagram of $\rho^{n}$(t) for a set of sequences $y_{0},y_{1}\cdots,y_{T}$: The red line represents the Kraus operator $K_{y}$ in $\{K\}$, the black line represents the Kraus operator $R_{y}$ in $\{R\}$, and the blue line represents the Kraus operator $A_{y}$ in $\{A\}$}
	\label{network}
\end{figure}
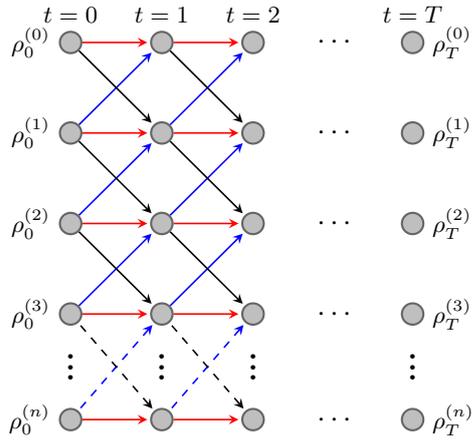

To calculate the parameters of the SHQMM, assuming that a set of sequences $y_{0},y_{1}\cdots,y_{T}$ are known, the conditional density matrix evolution under the measurement result $y_{i}$ is shown in Fig.\ref{network} based on transport system. It can be seen that Fig.\ref{network} is similar to a neural network and shows the process of forward propagation through time $t$. This suggests that SHQMM is promising as a physical realization pathway for QNN, which will be our further research. This demonstrates the connection of quantum state evolution among the different subspaces $n$ in QCME and the conversion relationship among the probabilities $Tr(\rho^{(n)})$. Compared to previous work on HQMM (the probabilities for the measurement value $y_{i}$ depend on $\rho=\sum_{n}\rho^{(n)}$), the property illustrated in Fig.\ref{network} also displays the contribution variance of different $\rho^{(n)}$ to the probabilities of obtaining the measurement value $y_{i}$ at time $t_{i}$. Therefore, our model produces a more stable and robust model structure (as seen in experimental results).

Here, based on the QCME (Eq.\ref{eq14}), we can write a probability function using the following equations.

\begin{widetext}
\begin{equation}\label{model}
	\begin{aligned}
		\rho_{T}^{(0)}&=K_{y_{T-1}}\rho_{T-1}^{(0)}K_{y_{T-1}}^{\dagger}+A_{y_{T-1}}\rho_{T-1}^{(1)}A^{\dagger}_{y_{T-1}},\\
		\rho_{T}^{(1)}&=R_{y_{T-1}}\rho_{T-1}^{(0)}R^{\dagger}_{y_{T-1}}+K_{y_{T-1}}\rho_{T-1}^{(1)}K_{y_{T-1}}+A_{y_{T-1}}\rho_{T-1}^{(2)}A^{\dagger}_{y_{T-1}},\\
		&\qquad \qquad \qquad \qquad \dots\\
		\rho_{T}^{(n)}&=R_{y_{T-1}}\rho_{T-1}^{(n-1)}R^{\dagger}_{y_{T-1}}+K_{y_{T-1}}\rho_{T-1}^{(n)}K_{y_{T-1}}+A_{y_{T-1}}\rho_{T-1}^{(n+1)}A^{\dagger}_{y_{T-1}},\\
		&\qquad \qquad \qquad \qquad \dots\\
		\rho_{T}^{(N_{max})}&=R_{y_{T-1}}\rho_{T-1}^{(N_{max}-1)}R^{\dagger}_{y_{T-1}}+K_{y_{T-1}}\rho_{T-1}^{(N_{max})}K_{y_{T-1}}.
	\end{aligned}
\end{equation}
\end{widetext}

where $K_{i,\mu}$, $K_{3,\mu}$ and  $K_{4,\mu}$ of Eq.\ref{eq14} denote $K_{y_{i}}$, $R_{y_{i}}$ and $A_{y_{i}}$, respectively. $N_{max}$ denotes the maximum value of $n$. The probability of $y_{i}$ is $P(y_{i})=\sum_{n}{\rm Tr}[K_{y_{i}}\rho^{(n)}(t)K_{y_{i}}^{\dagger} + R_{y_{i}}\rho^{(n-1)}(t)R_{y_{i}}^{\dagger} + A_{y_{i}}\rho^{(n+1)}(t)A_{y_{i}}^{\dagger}]$. 

From Eq.\ref{model}, the probability of sequences $y_{0},y_{1}\cdots,y_{T}$ can be easily obtained

\begin{equation}\label{PS}
	\begin{aligned}
		P_{y_{0},y_{1}\cdots,y_{T}}&=Tr(\rho_{T})\\
		\rho_{T}&=\rho^{(0)}_{T}+\rho^{(1)}_{T}+\cdots+\rho^{(...)}_{T}.\\  
	\end{aligned}
\end{equation}

To compute the parameters of the SHQMM, we propose a maximum likelihood estimation method, based on the results in \cite{Siddarth,Sandesh}. Firstly, we use the probability function to derive all possible Kraus operators, and then use the gradient descent algorithm to find the matrix form of the Kraus operator that satisfies the minimum probability function of the given sequence. This turns parameter-solving into an optimization problem.

\begin{equation}\label{optimize_problem}
	\begin{aligned}
		&{\rm minimize}_{\{K, R, A\}} \quad \mathscr{L}(\{K, R, A\})=-{\rm ln}\, {\rm Tr}(\rho_{T})  \\
		&{\rm subject \ to} \quad\quad\quad \sum_{y\in  O}K_{y}^{\dagger}K_{y}+R_{y}^{\dagger}R_{y} + A_{y}^{\dagger}A_{y}=I.
	\end{aligned}
\end{equation}

Stack $K_{y}$, $R_{y}$, $A_{y}$ by column to form a new matrix $\kappa=[K_{1},K_{2},\cdots,R_{1},R_{2},\cdots,A_{1},A_{2},\cdots]$ with dimension $3 {\rm  dim} O\cdot m\times m$($m$ is the dimension of the Kraus operator)  . So the constraint condition in Eq.\ref{optimize_problem} can be rewritten as

\begin{equation}
	\kappa^{\dagger}\kappa = I, \kappa \in \mathbb{C}^{3{\rm dim}O\cdot m\times m}.  
\end{equation}

We summarize all the above steps into an algorithm for solving the Kraus operator. The specific steps are shown in the Algorithm\ref{alogrithm1}.

\begin{algorithm}[H]  
	\caption{Leaning SHQMM using gradient descent method on Stiefel manifold}  
	\label{alogrithm1}  
	\begin{algorithmic}[1]  
		\Require  
		Training data $D\in \mathbb{N}^{M\times l}$ , $W$ is number of sequences and $l$ is the length of sequence.
		\Ensure  $\{\mathbf{K_{i}}\}_{i=1}^{{\rm dim}O}$, $\{\mathbf{R_{i}}\}_{i=1}^{{\rm dim}O}$, $\{\mathbf{A_{i}}\}_{i=1}^{{\rm dim}O}$$\cdots$
		\State {\bf Initialize}: Complex orthogonal matrix on Stiefel ${\bf \kappa}\in \mathbb{C}^{3{\rm dim O}\cdot m\times m}$ and $\rho^{(0)}, \rho^{(1)} , \rho^{(2)}, \cdots, \rho^{(N-1)}$ and require that $\rho^{(i)}$ is positive semi-definite and $\sum_{i=0}^{N-1}\rho^{(i)}=\rho_{total}$, $\rho_{total}$ is density matrix.
		\label{code:fram:extract}  
		\State {\bf for} $epoch = 1:E$ {\bf do}
		\label{code:fram:trainbase}  
		\State   \quad split the data $D$ into B batches ${D_{B}}$
		\label{code:fram:add}  
		\State \quad {\bf for } $batch = 1:B$ {\bf do}
		\label{code:fram:classify}  
		\State \quad\quad Compute gradient $G^{\{K\}}_{i} = \frac{\partial\mathscr{L}}{\partial K_{i}^{*}}$, $G^{\{R\}}_{i} = \frac{\partial\mathscr{L}}{\partial R_{i}^{*}}$, $G^{\{A\}}_{i} = \frac{\partial\mathscr{L}}{\partial A_{i}^{*}}$ 
		\State \quad \quad Compute the like-hood function $\mathscr{L}$
		\State \quad \quad Stack $G_{i}^{\{K\}}$, $G_{i}^{\{R\}}$, $G_{i}^{\{A\}}$ vertically to construct $G=[G_{1}^{K},\cdots,G_{O}^{A}]^{T}$
		\State \quad \quad Construct $\mathbf{U} = [G|\kappa], \mathbf{V}=[\kappa|-G]$
		\State \quad \quad $G = \beta G_{old} + (1-\beta) G$
		\State \quad \quad Update $\kappa = \kappa-\tau \mathbf{U}(I+\frac{\tau}{2}\mathbf{V}^{\dagger}\mathbf{U})^{-1}\mathbf{V}^{\dagger}\kappa$
		\State \quad {\bf end for}
		\State  \quad Update learning rate $\tau = \alpha\tau$
		\State {\bf end for}
		\State Compute the DA function by using the value of probability function $\mathscr{L}$ \\
		\Return ${\{\mathbf{K}_{i}\}}_{i=1}^{{\rm dim}O}, {\{\mathbf{R}_{i}\}}_{i=1}^{{\rm dim}O}, {\{\mathbf{A}_{i}\}}_{i=1}^{{\rm dim}O}$ and DA
	\end{algorithmic}  
\end{algorithm}

In Algorithm\ref{alogrithm1}, $\tau$(learning rate), $\alpha$(decay factor) and $\beta$(momentum parameter) are the hyper-parameters, and DA is a function that describes the quality of the model.

As one of the core evaluation metrics of HQMM, DA \cite{zhao2010norm} is able to measure the performance among models with different structures and functions in a relatively fair way \cite{adhikary2019learning}, defined as: 

\begin{equation}
	DA = f(1+\frac{{\rm log}_{\iota} P(D|M)}{l}).
\end{equation}

Where, $D$ is data, $M$ is model. $l$ is the length of the sequence, and $\iota$ is the number of output symbol in the sequence. The function $f(\cdot)$ is a non-linear segmented function that can map any argument in $(-\infty,1]$ to $(-1,1]$, defined as:

\begin{equation}
	f(x)=\left\{ 
	\begin{aligned}
		&x, & x\geq 0,\\
		&\frac{1-e^{-0.25x}}{1+e^{-0.25x}}, & x<0.
	\end{aligned}
	\right.
\end{equation}

The model can perfectly predict the Markov sequence if $DA=1$ and the model better than random model for $DA>0$. 

In the SHQMM, different models can produce different prediction effects for the same sequences, depending on the number $\mathcal{M}{q}$ of initialized conditional density matrices and the connections between them. Eq.\ref{model} describes the closest connection between the conditional density matrix and the number $\mathcal{M}{q}$, which is equal to $n$.

 The optimal solution of HQMM is one of the optimal solutions of Eq.\ref{optimize_problem}. Therefore, periodic boundary conditions are applied to the first and last conditional density matrices, similar to the arrangement of atoms in a crystal. 

\begin{equation}\label{boundary conditions}
	\begin{aligned}
		\rho_{T}^{(0)}&=K_{y_{T-1}}\rho_{T-1}^{(0)}K_{y_{T-1}}^{\dagger}+A_{y_{T-1}}\rho_{T-1}^{(1)}A^{\dagger}_{y_{T-1}}\\&+R_{y_{T-1}}\rho_{T-1}^{(N_{max})}R^{\dagger}_{y_{T-1}},\\
		\rho_{T}^{(N_{max})}&=R_{y_{T-1}}\rho_{T-1}^{(N_{max}-1)}R^{\dagger}+K_{y_{T-1}}\rho_{T-1}^{(N_{max})}K_{y_{T-1}}\\&+A_{y_{T-1}}\rho_{T-1}^{(0)}A^{\dagger}_{y_{T-1}}.
	\end{aligned}
\end{equation}

Different from Eq.\ref{model}, Eq.\ref{boundary conditions} includes periodic boundary conditions. This empirical improvement is due to the fact that the Kraus operators $(R, A)$ often converge to zero during the optimization process, which means that the model has essentially degenerated into a HQMM. This improvement ensures the stability of the learning process.

More detailed cases are presented in supplemental material.

\paragraph{Extend the expandability ability of the SHQMM}

If we need to further improve the complexity of the model, we can set the parameters $N_{max}=4, 5, 6, \cdots$ and apply a more complicated connection defined as $k$-local. Thus, the general SHQMM can be defined as a tuple $\lambda_{SQ}=(\mathbb{C}^{m}, k{\rm-local}, {K_{y}^{j}}, {y\in O}, {j=2k+1}, {\rho_{0}^{(i)}}, i\in [0, {N_{max}-1}])$ with the following conditions:\\

	(1) $\rho_{0}^{(i)}$ is conditional density matrix, ${\rm Tr}(\sum_{i=0}^{N_{max}-1}\rho^{(i)}_{0})=1$.\\
	
	(2) For every Kraus operator, $K_{y}^{j}:\mathbb{C}^{m}\to \mathbb{C}^{m}$ and $\sum_{y,j}\left (K_{y}^{j}\right )^{\dagger}K_{y}^{j}=I$.\\
	
	(3) The evolution of $\rho^{(i)}$ follows Eq.\ref{evolution_shqmm}.
	
	 \begin{equation}\label{evolution_shqmm}
	 	\begin{aligned}
		&\rho^{(i)}(t+\Delta t)=\\
		&\sum_{j^{\prime}=-\lfloor\frac{j}{2}\rfloor}^{\lfloor\frac{j}{2}\rfloor}\sum_{y\in O}K_{y}^{j^{\prime}+1+\lfloor\frac{j}{2}\rfloor}\rho^{(i-j^{\prime})}(t)K_{y}^{j^{\prime}+1+\lfloor\frac{j}{2}\rfloor \dagger},
		\end{aligned}
	\end{equation}

where the periodic boundary conditions should be applied and conditional density matrix beyond index range should be zeroed, that is, $0 \leq i-j'\leq N_{max}-1$. $\lfloor\frac{j}{2}\rfloor$ equal to $k$, $j'\in \{-\lfloor\frac{j}{2}\rfloor,-\lfloor\frac{j}{2}\rfloor+1$,$\dots$,$\lfloor\frac{j}{2}\rfloor-1,\lfloor\frac{j}{2}\rfloor\}$.\\

(4) The probability of observation symbol $y$ is
\begin{equation}\label{Probability_General_shqmm}
	 	\begin{aligned}
	&p(y)=\\
	&\sum_{i=0}^{N-1}{\rm Tr}(\sum_{j^{\prime}=-\lfloor\frac{j}{2}\rfloor}^{\lfloor\frac{j}{2}\rfloor}K_{y}^{j^{\prime}+1+\lfloor\frac{j}{2}\rfloor}\rho^{(i-j^{\prime})}(t)K_{y}^{j^{\prime}+1+\lfloor\frac{j}{2}\rfloor \dagger}),
	 	\end{aligned}
\end{equation}

where, $k$-local represents the relationship between different conditional density matrix and $j$ represents the number of Kraus operator classes. 

In summary, enhancing state correlations and moving beyond simple neighbor connections enhances SHQMM's applicability to diverse time series problems.

\paragraph{Comparison of properties of SHQMM and HQMM}

We use a simple case to illustrate the correlation and difference between SHQMM and HQMM. For 1-local model, by summing the conditional density matrix $\rho^{(n)}$ with index $n$ in Eq.\ref{model}, we obtain:

	\begin{equation}\label{After_summing_simple_case}
		\begin{aligned}
			&\rho(t+\Delta t)\\
			&=\sum_{y}K_{y}\rho(t) K_{y}^{\dagger} + \sum_{y} R_{y}\rho(t)R_{y}^{\dagger} + \sum_{y} A_{y}\rho(t)A_{y}^{\dagger}\\
			&=\sum_{\omega_{y}}K_{\omega_{y}}\rho(t)K_{\omega_{y}}^{\dagger}.
		\end{aligned}
	\end{equation}

The second equal sign in Eq.\ref{After_summing_simple_case} shows a formal relationship between SHQMM and HQMM, but our model has clear physical implications compared to the auxiliary dimension $\omega_{y}$ in Eq.\ref{eq5} (HQMM with $\omega_y=3$). This indicates that the SHQMM is a valid HQMM at the same time.

The differences between SHQMM and HQMM lie in the following aspects:

\begin{itemize}
	\item SHQMM makes density matrix have a hierarchy structure as shown in Fig.\ref{network}, and the density matrix evolves through multiple channels.
	\item Kraus operator$\{K_{y}^{j}\}_{y\in O}^{j=2k+1}$ acts on different conditional density matrix $\rho^{(n)}$ in SHQMM, whereas in HQMM, Kraus oprator $\{K_{\omega_{y}}\}_{y\in O}$ acts on total density matrix $\rho$.
	\item SHQMM can be derived from actual physical systems, such as quantum transport systems, as shown in Fig.\ref{network}.
\end{itemize}

The SHQMM can reflect the relationship between hidden states and is more suitable for handling more complex data than the HQMM. From a physical system implementation point of view, the quantum conditional master equation may differ from the quantum transport system for other open quantum systems\cite{Yoshitaka,Ishizaki,Jin}, resulting in different split hidden quantum Markov models. The Bayesian rule for SHQMM is:

\begin{widetext}
	\begin{equation}\label{conditional probability}
		\rho^{(n)}_{y|x}(t+ \Delta t)=\frac{\sum_{j^{\prime}=\lfloor\frac{j}{2}\rfloor} ^{\lfloor\frac{j}{2}\rfloor}K_{y}^{j^{\prime}+1+\lfloor\frac{j}{2}\rfloor}\rho^{(n-j^{\prime})}(t)K_{y}^{j^{\prime}+1+\lfloor\frac{j}{2}\rfloor\dagger}}{{\rm Tr}\left( \sum_{n=0}^{n=N-1}\sum_{j^{\prime}=\lfloor\frac{j}{2}\rfloor} ^{\lfloor\frac{j}{2}\rfloor}K_{y}^{j^{\prime}+1+\lfloor\frac{j}{2}\rfloor}\rho^{(n-j^{\prime})}(t)K_{y}^{j^{\prime}+1+\lfloor\frac{j}{2}\rfloor\dagger}
			\right)}.
	\end{equation}
\end{widetext}

Eq.\ref{conditional probability} expresses, under the principle of conditional probability, the quantum state corresponding to the system when the system is observed as $x$ at $t$ and as $y$ at $t+ \Delta t$. Table \ref{Compared_hqmm_shqmm} shows the properties of the SHQMM and HQMM.

\begin{table*}
	\caption{The properties of HQMM and sHQMM}
	\label{Compared_hqmm_shqmm}
	\begin{tabular}{cccccc}
		%{p{2cm}p{1.5cm}p{4cm}p{3.5cm}p{3cm}p{2cm}}
		\toprule
		Model &State&	\makecell[c]{Transition \\and Emission}&Probability&Bayesian Rule&Evolution  \\ \hline
		SHQMM &$\{\rho^{(i)}\}_{i=0}^{N-1}$&\makecell[c]{quantum channel\\ $\mathcal{K}_{s}$}&Eq.\ref{Probability_General_shqmm}&$\rho_{y|x}^{(n)} $ & Eq.\ref{evolution_shqmm}\\
		\midrule
		HQMM & $\rho$&\makecell[c]{quantum channel\\$\mathcal{K}$}&${\rm Tr}(\sum_{\omega_{y}}K_{\omega_{y}}\rho(t)K^{\dagger}_{\omega_{y}})$&$ Eq.\ref{eq5}$&$\sum_{\omega_{y}}K_{\omega_{y}}\rho(t)K_{\omega_{y}}^{\dagger}$\\
		\bottomrule
	\end{tabular}
\end{table*}

\section{Experiment and results \label{Experiment and Results}}

In this section, we applied quantum and classical data to train and test our SHQMM. All experiments were performed on an experimental platform outfitted with Intel Core i7 CPU and 16 GB RAM, trials were executed on Python 3.8 with PyTorch and qiskit libraries for modeling and evaluating. The source code and further details are available at \href{https://github.com/Aderson10086/Hidden-Quantum-Markov-model-and-its-extensions}{here}.

\paragraph{Quantum data}
Firstly, we used the quantum data generated by a quantum mechanical process in Ref.\cite{Siddarth}.  The quantum data has six hidden states and six observational values. The size of the quantum data is $40\times 3000$. 

\begin{figure}[b]
	\centering
	\subfigbottomskip=2pt %两行子图之间的行间距
	\subfigcapskip=-5pt %设置子图与子标题之间的距离
	\subfigure[]{
		\includegraphics[scale=0.65,trim={1.2cm 0 0 0}]{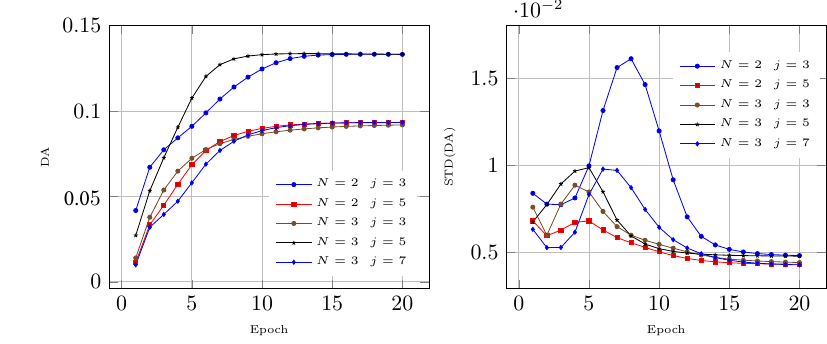}}
	\subfigure[]{
		\includegraphics[scale=0.65,trim={1.2cm 0 0 0}]{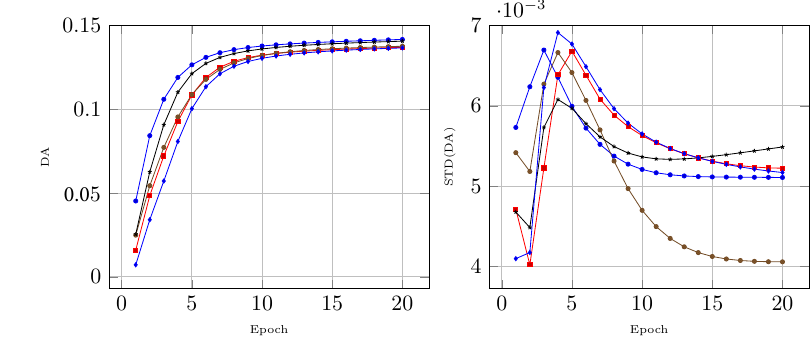}}
	\subfigure[]{
		\includegraphics[scale=0.65,trim={1.2cm 0 0 0}]{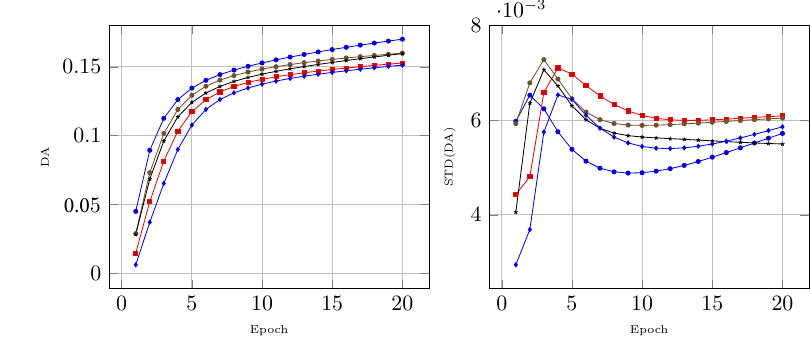}}
	\caption{The training result of the different SHQMM for quantum data under different parameters $N$ and $j=2k+1$. The subfigures (a), (b), (c) represent the training results for choosing single, double and three qubits quantum system, respectively.}
	\label{quantum_figure}
\end{figure}

\paragraph{Training and validation}
We use $20\times 3000$ data to train our model, generating a total of 20 models. Simultaneously, we used $10\times 3000$ data to verify the model, and the remaining data were used to test the model. The results are shown in Fig.\ref{quantum_figure} with hyperparameters $\tau=0.95, \alpha=0.95, \beta=0.90$ (More calculation results can be found in supplemental material). Fig.\ref{quantum_figure} shows several models $\lambda_{SQ}$ for single, double and three qubits quantum system which are used to construct the SHQMM under the different parameters $N$ and $k$. It found : (1) With the increase of the qubit number, the value of DA also increases and reaches a stable value at about $20$ epochs (three qubits exceed $20$ epochs). (2) Apart from the single bit, the connection mode (different k-local) and $N$ have little effect on the DA value. (3) While a higher number of quantum bits results in a relatively larger value of DA, the standard deviation (STD) decreases as the number of qubits increases during evaluation with alternative data. This suggesting potential overfitting.

To further test the reliability of our model, we will evaluate it from several perspectives.

\paragraph{Initialize Kraus}
In Ref. \cite{Sandesh}, it was stated that the training outcome of HQMM is susceptible to the initial Kraus operators in smaller models. Thus, this study investigates the effect of the initial position of Kraus operators on Stiefel manifolds for SHQMM. Eq. \ref{Dis_Manifolds} is utilized to evaluate the distance between various initial positions.

\begin{equation}\label{Dis_Manifolds}
	D(\kappa_{1}, \kappa_{2})=||\kappa_{1}\kappa_{2}^{\dagger}-I||_{2}.
\end{equation}

\begin{figure}[b]	
	\centering
	\includegraphics[scale=0.75,trim={0.6cm 0 0 0},clip]{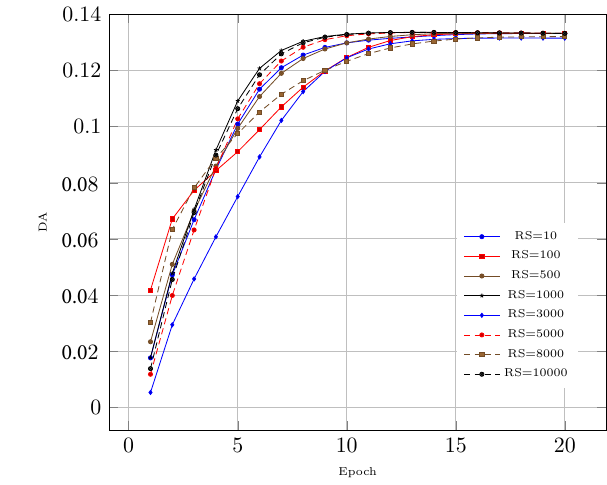}
	\caption{The training resulut of SHQMM($N_{max}=3$, AR, Single Qubit) in random initialization}
	\label{Initial_Kraus}
\end{figure}

When $\kappa_{1}=\kappa_{2}$, $D=0$. The initialization method for the Kraus operator is presented in Algorithm \ref{Initialize_Method}. The varying behaviors of the DA are depicted in Figure \ref{Initial_Kraus} for different random initialization seeds (RS). Results clearly shows that our proposed SHQMM remains stable regardless of the initial Kraus values, thus confirming its validity.

\begin{algorithm}[H]
	\caption{Initial Kraus operator on Stiefel manifolds}
	\label{Initialize_Method}
	\begin{algorithmic}[1]
		\Require
		the dimension of Kraus operator $m$, the class of Kraus operator $j$, the dimension of observable space ${\rm dim}O$ and random seed RS
		\Ensure
		Kraus operator $\{K_{y}^{j}\}$
		\State{\bf Initialize}: random vector $\vec{v}_{1}={\rm random}(m\cdot j \cdot {\rm dim}O, 1, {\rm RS})$, zero matrix $\kappa={\rm zeros}(m\cdot j\cdot {\rm dim}O, 2m)$, $\kappa(:,1)=\frac{\vec{v}_{1}}{||\vec{v}_{1}||}$
		\State{\bf for} $s=2:2m$ {\bf do}
		\State \quad $\vec{v}_{s}={\rm random}(m\cdot j\cdot {\rm dim O},1)$
		\State \quad {\bf if} all column vectors in $\kappa$ and $\vec{v}_{s}$ are orthogonal {\bf do}
		\State \quad \quad $\vec{v}_{s}=\frac{\vec{v}_{s}}{||\vec{v}_{s}||}$; $\kappa(:,s)=\vec{v}_{s}$
		\State \quad {\bf else} {\bf do} 
		\State \quad \quad  Schmidt Orthogonalization of $\kappa$ and $\vec{v}_{s}$
		\State \quad \quad $\vec{v}_{s}=\frac{\vec{v}_{s}}{||\vec{v}_{s}||}$; $\kappa(:,s)=\vec{v}_{s}$
		\State \quad {\bf end if}
		\State {\bf end for}
		\State construct new matrix $\kappa^{\prime}=\kappa(:,1:m)/\sqrt{2}+\kappa(:,m+1:2m)/\sqrt{2}$
		\State split $\kappa^{\prime}$ into Kraus operator
	\end{algorithmic}
\end{algorithm}

\paragraph{Selection of effective models}

Given the various methods available to construct models, selecting the most effective one for a given dataset is a critical challenge. Typically, the expressiveness of a model is directly linked to the number of its parameters. The number of parameters for SHQMM is as follows:

\begin{equation}
	\mathscr{N}_{P}=m^{2}\cdot j.
\end{equation}

\begin{figure}[htb]	
	\centering
	\subfigure[]{
		\includegraphics[scale=1,trim={1.2cm 0 0 0}]{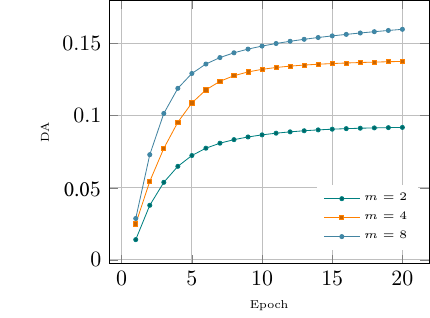}}
	\subfigure[]{
		\includegraphics[scale=1,trim={1.2cm 0 0 0}]{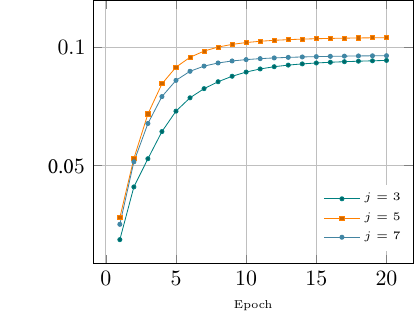}}
	\caption{The relationship between the training outcome of SHQMM and the number of parameters. (a) represents the variation of DA with the dimension $m$ of the Kraus operator. (b) represents the variation of DA with the $j$ of the conditional density matrix. The training results are more sensitive to $m$ than $j$}
	\label{DA_Variational}
\end{figure}

The corresponding results are shown in Fig.\ref{DA_Variational} 
to obtain best training results, we should change the dimension $m$ of Kraus operator firstly and then adjust the parameter $j$ for a given sequence.

\paragraph{Hyperparameters selection for the model}

\begin{figure}[b]	
	\centering
	\includegraphics[scale=0.7,trim={1.2cm 0 0 0}]{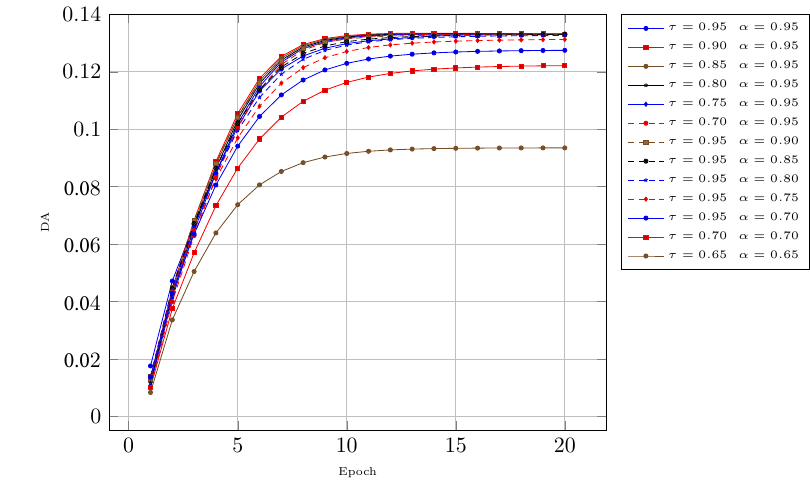}
	\caption{The effect of DA on different hyperparameters tuple($\tau, \alpha$)($\beta=0.90$).The best DA is 0.1332($\tau=0.95,\alpha=0.95$) and the worst DA is 0.0951($\tau=0.65, \alpha=0.65$)}
	\label{Hyper_alpha_tau}
\end{figure}

\begin{figure}[b]
	\centering
	\subfigure[]{
		\includegraphics[scale=0.6,trim={2cm 0 0 0}]{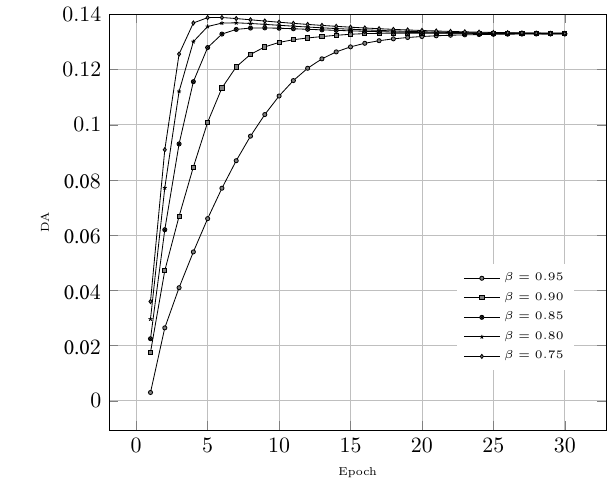}}
	\subfigure[]{
		\includegraphics[scale=0.6,trim={2cm 0 0 0}]{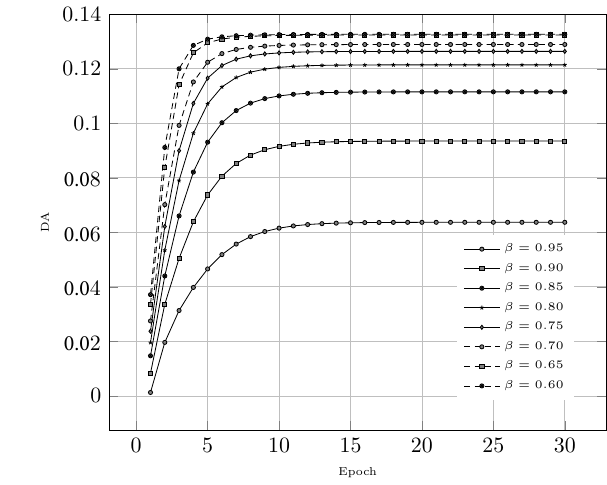}}
	\caption{Choose the hyperparameters for sHQMM. (a) shows the DA of  $\tau=0.95,\alpha=0.95$;(b) shows the DA of $ \tau=0.65,\alpha=0.65$.}
	\label{Momentum_Hyperparameters}
\end{figure}

To obtain the optimal $DA$, Algorithm \ref{alogrithm1} employs three hyperparameters, namely $\tau$, $\alpha$, and $\beta$. Fig. \ref{Hyper_alpha_tau} demonstrates the impact of varying hyperparameters on $DA$, indicating that $DA$ is more sensitive to changes in $\alpha$ as compared to $\tau$. Moreover, the existence of multiple local optima in SHQMM is evident. To identify the global optimum, we investigated the effect of momentum parameter $\beta$ on $DA$ for the best case ($\tau=0.95,\alpha=0.95$) and the worst case ($\tau=0.65,\alpha=0.65$), as presented in Fig.\ref{Momentum_Hyperparameters}. It was observed that $DA$ may reach the global optimum at $\tau=0.95$, $\alpha=0.95$, and $\beta=0.90$, and that $DA$ can be further enhanced by selecting a different $\beta$. However, after computing the distance of Kraus solutions between different hyperparameters ($\tau=0.95$, $\alpha=0.95$, $\beta=0.90$ and $\tau=0.65$, $\alpha=0.65$, $\beta=0.60$) using Eq. \ref{Dis_Manifolds}, we discovered that their $DA$ values were comparable despite locating at different positions on the Stiefel manifold.

\paragraph{Classical data}

\begin{figure}[h]
	\centering
	\subfigure[]{
		\includegraphics[scale=0.6,trim={1.5cm 0 0 0}]{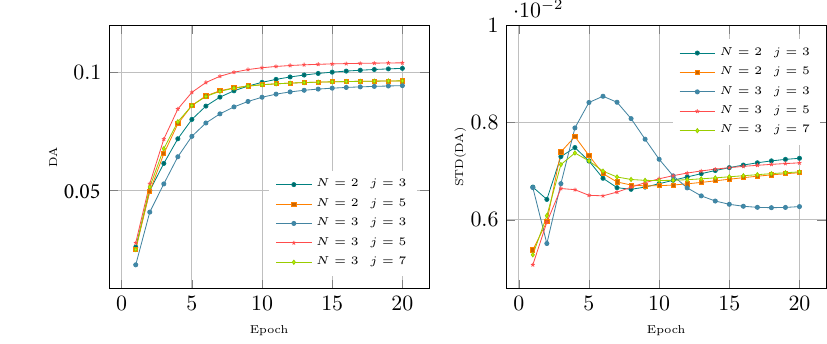}}
	\subfigure[]{
		\includegraphics[scale=0.6,trim={1.5cm 0 0 0}]{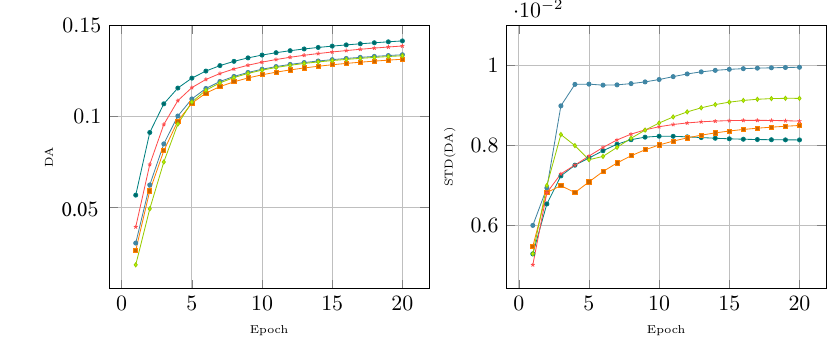}}
	\subfigure[]{
 		\includegraphics[scale=0.6,trim={1.5cm 0 0 0}]{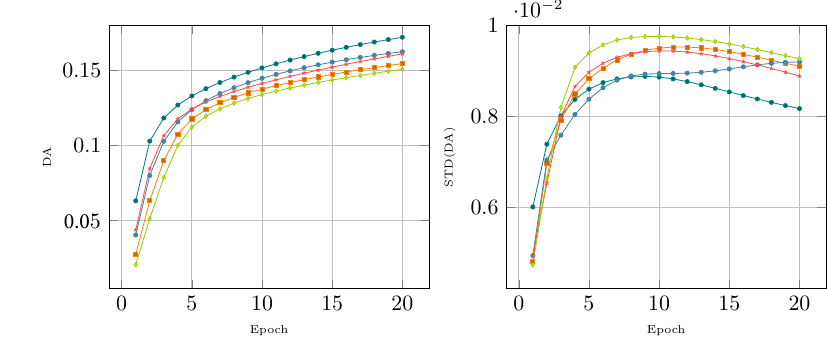}}
	\caption{The training results of SHQMM for classical data under different parameters $N$ and $j=2k+1$. The subfigures (a), (b), (c) represent the training results for choosing single, double and three qubits quantum system, respectively.}
	\label{classical_figure}
\end{figure}

To conduct a thorough evaluation of the model, classical data generated by a hidden Markov process with transition matrix ${\bf T}$ and emission matrix ${\bf{C}}$ were utilized to compute the Kraus operator and determine $DA$. The results obtained from the classical data are presented in Fig. \ref{classical_figure}, where hyperparameters were set to $\tau=0.95$, $\alpha=0.95$, and $\beta=0.90$.

Similar to the quantum case, the value of $DA$ also increases and reaches a stable state after approximately 20 epochs (for three qubits, it took 20 epochs to stabilize). The different values of k-local have little impact on the $DA$ value, and the standard deviation (STD) continues to decrease as the qubit number increases for the testing data, possibly due to model overfitting. Additional test results can be found in supplemental material.

\begin{equation*}
	\begin{aligned}
		\mathbf{T} =& 
		\begin{pmatrix}
			0.8 & 0.01&0&0.1&0.3&0\\
			0.02&0.02&0.1&0.15&0.05&0\\
			0.08&0.03&0.1&0.4&0.05&0.5\\
			0.05&0.04&0.5&0.35&0 &0.5\\
			0.03.&0.5&0.03&0&0.6&0\\
			0.02&0.4&0.27&0&0&0
		\end{pmatrix},\\
		\mathbf{C} =& \begin{pmatrix}
			0.2 & 0 & 0.05 & 0.95 &0.01 & 0.05\\
			0.7 & 0.1 & 0.05 &0.01& 0.05& 0.05\\
			0.05&0.8&0.1&0.02&0.05&0.04\\
			0.04&0.04&0.02&0&0.84&0.11\\
			0.01&0.03&0.7&0.01&0.02&0.2&\\
			0&0.03&0.08&0.01&0.03&0.55
		\end{pmatrix}.
	\end{aligned}
\end{equation*}

\paragraph{Result}

The numerical experimental results of SHQMM on quantum and classical data are displayed in supplemental material. The current results are likely to be overfitted given the complexity of the relationship between complex model and simple data with lower number and dimensionality of hidden states \cite{elliott2021memory}. Keeping the hyperparameters constant, the training step converges more and more slowly as the number of qubits increases. 

SHQMM exhibits good robustness in numerical experiments and maintains stable performance in multiple task scenarios. For different sequences, the model maintains ${\rm STD}(DA)<0.01$. Different initial values of the Kraus operator have a negligible effect on the convergence of the model, due to the hierarchical structure of the conditional density matrix.

\section{Analysis and discussion \label{discussion} }

In this section, the performance of SHQMM is analyzed from both experimental and theoretical perspectives.

\paragraph{Performance comparisons}

It is imperative to acknowledge the existence of structural differences between different models. Remarkably, the model presented here represents a groundbreaking development in our recognized field, as it is uniquely capable of articulating a meticulous representation of the internal state of a quantum system. Consequently, the task of identifying models that are precisely aligned to the same baseline for performance evaluation becomes a formidable challenge. To address this, we have wisely chosen to compare models that exhibit maximum congruence in both structure and functionality. Table. \ref{compares} compares the DA of quantum data obtained by the SHQMM, HMM and HQMM in \cite{Siddarth} as follows:

\begin{table}[H]
	\caption{Comparison of HMM, HQMM and SHQMM on DA. The presentation of complete data under different parameters is shown in \cite{Siddarth} and supplemental material.}
	\label{compares}
	\centering 
	\renewcommand
	\arraystretch{1}
	\tabcolsep=0.6cm
	\begin{tabular}{cc}%{\textwidth}
		\toprule
		Model & DA\\
		\midrule
		$2,6-$HMM($L$) & 0.0327\\
		$2,6,1-$HQMM ($L$) & 0.1303\\
		$3,$AR-SHQMM & $\geq$0.1700\\
		\bottomrule
	\end{tabular}
	%\vspace{\baselineskip}
\end{table}

$2,2$-HMM($L$) stands for the classical Hidden Markov model with two hidden states and two observables, and $L$ stands for the fact that the value is obtained by learning. Similarly, $2,2,1$-HQMM($L$) denotes the quantum hidden Markov model having two hidden states, two observables, with one Kraus operator per observable. For SHQMM, $3,$AR means that the number of hidden states is $3$, and $AR$ means that the current model has $3$ hidden states and connectivity is adjacent. $\geq$ indicates that the value did not converge at the number
of iterations set.

 As can be seen from Table. \ref{compares}, SHQMM improves DA by about $23.35\%$ over the HQMM algorithm, while the quantum versions of the algorithm all have better performance in model quality over the classical algorithm. 

\paragraph{Complexity analysis}

Complexity serves as a theoretical measure encompassing both the computational and time costs associated with an algorithm. To ensure consistency between various models and their original literature in terms of description, we have retained their respective symbol systems in this chapter. The complexity of HMM, HQMM and SHQMM is analyzed as follows:

\begin{itemize}
	
	\item Building upon previous research \cite{Siddarth}, a classical HMM algorithm can be succinctly characterized as $(n, s)$, where $n$ denotes the number of hidden states and $s$ represents the number of observations. In the context of an HMM with a prediction sequence of length $T$, the complexity arises as each state transition involves $n$ states undergoing transitions, with each state having $n$ possible transitions, resulting in a complexity denoted as $O(Tn^2)$\cite{westhead2017hidden}.

	\item Similarly, the HQMM is denoted by the ternary $(n, s, w)$. Introducing the parameter $w$, alongside $n$ and $s$, accounts for the number of Kraus operators per observable. In a manner akin to HMM, where a HQMM prediction sequence of length T is considered, there exist $n$ states undergoing transitions, with each state producing $w$ observations. Consequently, the complexity of HQMM is denoted as $O(Tnw)$. In scenarios of comparable scale, the complexity of HQMM is notably influenced by the parameter $w$, which can be independently set based on the application scenario. Generally, in smaller-scale tasks, $n < w$ may occur, while in longer sequence tasks, $n > w$ is more likely.

	\item The model structure of SHQMM is denoted as $(N_{max},s,k)$, where $N_{max}$ represents the number of hidden states in the model, equivalent to $n$ in other models. $k$ represents the connectivity state which is related to the number of Kraus operators. Due to structural constraints, $2k+1 \leq N_{max}$. Similarly to HMM, SHQMM exhibits parallels in complexity analysis. During the transition process among $N_{max}$ states, each state yields $2k+1$ observational outputs. Therefore, the complexity is expressed as $O(TN_{max}(2k+1))$. Notably, due to the constraints of the connection method, imposing an upper limit on the complexity of SHQMM, expressed as $O(TN_{max}(2k+1)) \leq O(TN_{max}^2)$.

\end{itemize}

\paragraph{Discussion of Quantum Advantage}

Exploring quantum advantages is a shared pursuit among all quantum machine learning algorithms.

In prior research, a substantial body of work has focused on examining the benefits of quantum versions of HMM. Noteworthy contributions, exemplified by \cite{Alex} and \cite{Siddarth}, have highlighted the advantages observed in numerical experiments of Quantum HMM (HQMM), particularly in the DA. Additionally, certain studies have undertaken theoretical analyses to elucidate these quantum advantages. Notably, \cite{LiuQing} and \cite{elliott2021memory} provide insights into the advantages of HQMM, emphasizing its physical performance and memory efficiency. Furthermore, \cite{markov2022implementation} illustrates the quantum approach's superiority in terms of state space complexity. Experiments in this work align with the aforementioned research, further substantiating the merits associated with quantum-based approaches.

According to the analysis of complexity, it is evident that HQMM and SHQMM demonstrate advantages when $n$ is larger. This implies that quantum solutions are more suitable for scenarios involving long sequence tasks. It should be acknowledged that in numerical experiments on classical computers, the results did not fully reflect this advantage. This is attributed to the additional computational cost incurred by the classical computer's quantum simulation process.

Nevertheless, the experimental results of our work, \cite{Siddarth} and \cite{Sandesh} still showcase the potential performance advantages of the quantum approach, namely the ability to convey more information with fewer hidden states. This high expressivity has been mathematically proven in \cite{markov2022implementation}, representing a potential advantage of quantum approaches. In terms of encoding efficiency, a substantial amount of information can be encoded onto a smaller number of quantum states, thereby enhancing the utilization of informational resources. This implies that SHQMM can be applied to large-scale tasks such as weather forecasting.

Furthermore, the superiority of SHQMM is underscored by its capacity to correspond to a tangible physical system, specifically the quantum transport system. This attribute imparts a lucid interpretability to SHQMM, enabling continuous trackigng of the internal intricacies of the quantum system throughout the evolutionary process further than HQMM. Also, owing to its association with an authentic quantum physical system, SHQMM demonstrates an increased affinity for quantum data, a trait supported by the results of numerical experiments. Together with other quantum versions of HMM models, SHQMM enriches the available selection of models for the study of quantum Markov processes.

It is noteworthy, however, that SHQMM's performance with some classical datasets appears suboptimal. Further analysis, informed by the insights of Michael \textit{et.al.}'s work \cite{kastoryano2023quantum} thereby constituting a focal point for our forthcoming research endeavors.

\section{Conclusion \label{conclusion} }
In this paper, the novel stochastic probabilistic graphical model, SHQMM, is introduced as our main theoretical contribution using quantum conditional master equations. An empirical improvement has also been implemented by introducing periodic boundary conditions to ensure the stability of the learning process.

Numerical experiments underscore that the model's performance on DA remains robust following the introduction of the new structure, displaying heightened insensitivity to the initial state and increased overall robustness.

SHQMM emerges as a valuable tool for elucidating the intricate relationships among hidden states within quantum systems, providing support and additional choices for addressing HMM challenges through quantum methods. The model's correspondence with quantum transport systems further enhances its appeal, offering promising prospects for physical implementation. Additionally, given the structural similarities, SHQMM assumes a pivotal role as a theoretical foundation for the physical realization of QNN.

\section{Acknowledgements \label{Acknowledgements}}
This work is supported by the Open Fund of Advanced Cryptography and System Security Key Laboratory of Sichuan Province (Grant No. SKLACSS-202210) and the National Key $\mathrm{R\&D}$ Program of China, Grant No.2018FYA0306703 Chengdu Innovation and Technology Project, No.2021-YF05-02413-GX and 2021-YF09-00114-GX, Sichuan Province key research and development project, No.2022YFG0315, Natural Science Foundation of Xinjiang Uygur Autonomous Region, Granted No. 2023D01A63.

\bibliography{Ref}
\bibliographystyle{quantum}

\end{document}